\documentclass[a4paper, oneside, twocolumn, notitlepage, 10pt]{extarticle_ecoc}
\usepackage{ecoc}
\usepackage{array}

\begin{document}
\selectlanguage{english}    
\captionsetup{singlelinecheck=false, justification=justified} 


\title{Geometric-Configuration Modulation: A Novel Free-Space Optical Communication Paradigm for $D/r_0\sim 5.0$ Turbulence Resistance}%


\author{
    Yu-Ming Bai\textsuperscript{(1,2)}, Ming-Han Ding\textsuperscript{(1,2)},
    Yu-Xuan Liu\textsuperscript{(1,2)} and Jun-Lin Li\textsuperscript{(1,2)}
}

\maketitle                  


\begin{strip}
    \begin{author_descr}

        \textsuperscript{(1)} Department of Physics, Tsinghua University, Beijing, China. 
        (\textcolor{blue}{\uline{center@mail.tsinghua.edu.cn}})

        \textsuperscript{(2)} State Key Laboratory of Low Dimensional Quantum Physics, Tsinghua University, Beijing, China.

    \end{author_descr}
\end{strip}


\begin{strip}
    \begin{ecoc_abstract}
        We propose Geometric-Configuration Modulation (GM), a novel AO-free FSO paradigm utilizing multi-source geometric configuration encoding and active correlative decoding. GM demonstrates exceptional resistance to strong atmospheric turbulence ($D/r_0\sim 5.0$) over a 1.2 m link in preliminary experiments. ©2026 The Author(s)
    \end{ecoc_abstract}
\end{strip}


\begin{figure*}[b!]
    \centering
    \vspace{-4mm}
    \includegraphics[width=\textwidth]{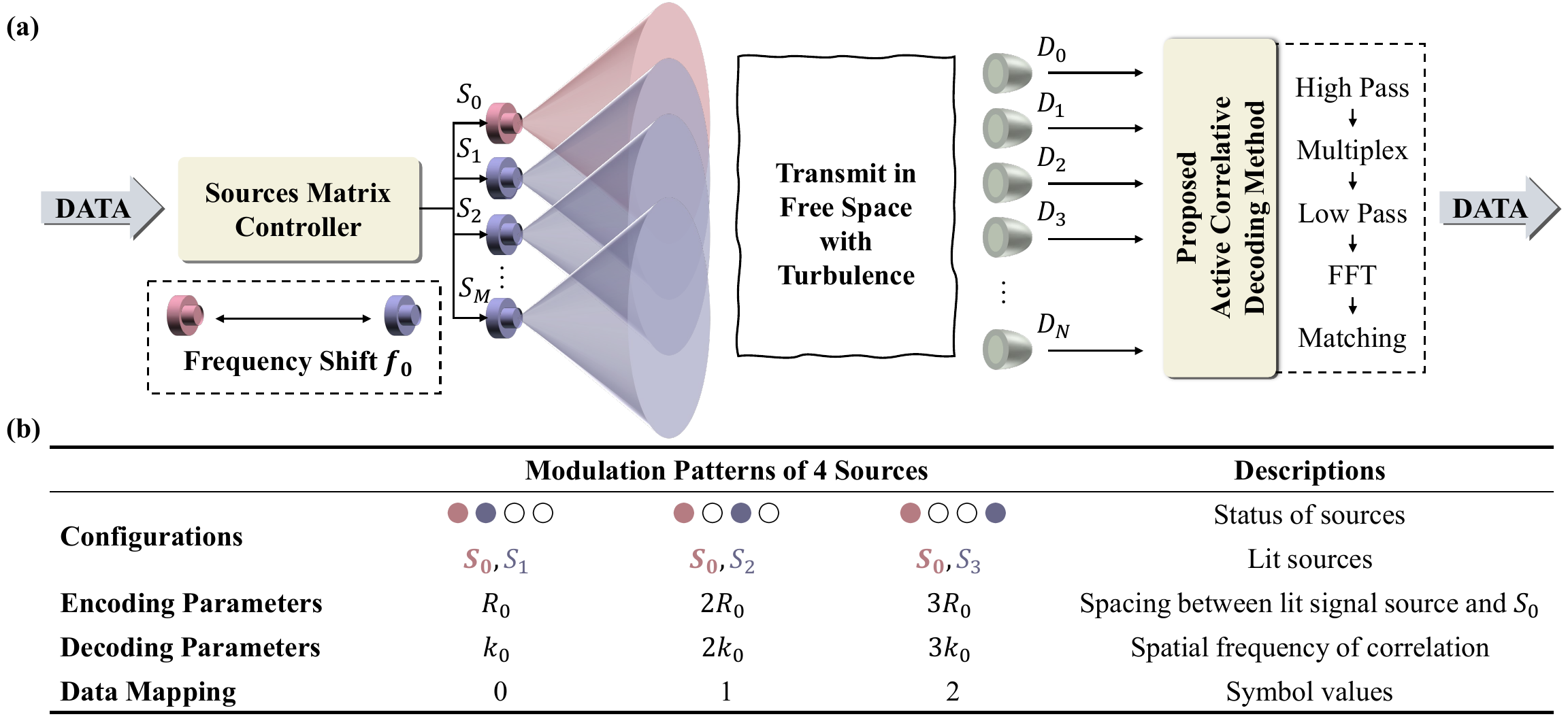}
    \caption{\textbf{Principle of GM-FSO.} \textbf{(a)}~System framework utilizing line-array sources. $S_0$: constantly-lit reference source; $S_1, \ldots, S_M$: signal sources; $D_0, D_1, \ldots, D_N$: detectors. \textbf{(b)}~Encoding/decoding example for a 4-source linear array, mapping geometric configurations to spatial frequencies of the correlation properties of the optical field in the overlapping region. $R_0$: source spacing; $k_0=2\pi R_0/\lambda L$: characteristic spatial frequency; $\lambda$: wavelength; $L$: source-to-detector distance.}
    \label{fig:01}
\end{figure*}

\section{Introduction}
Free-space optical (FSO) communication has emerged as a promising solution for satellite interconnects and terrestrial access owing to its high bandwidth and license-free spectrum \cite{RN1,RN2}. However, atmospheric turbulence induces intensity scintillation and wavefront distortion, severely degrading the signal-to-noise ratio (SNR) \cite{RN3,RN4}. 

Existing mitigation techniques—such as adaptive optics (AO) wavefront correction\cite{RN5,RN6}, spatial diversity\cite{RN7,RN8}, and optimized formats \cite{RN9,RN10,RN11}—primarily encode information in the first-order statistics (amplitude/phase) of the optical field, which are directly perturbed by turbulence. Consequently, performance improvements require significantly increased system complexity. A highly robust alternative is encoding information into the second-order spatial coherence\cite{RN13,RN14,RN15}, a principle rooted in Hanbury Brown and Twiss (HB-T) intensity interferometry \cite{RN12}. However, due to the long integration times of traditional correlation measurements, its application in high-speed FSO remains unexplored.

In this paper, we propose Geometric-Configuration Modulation (GM), a novel FSO paradigm that encodes information into the second-order coherence determined by the geometric configuration of light sources. By utilizing an active correlative decoding method via frequency shifting, GM extracts the second-order coherence to recover information with formidable, inherent resistance to turbulence. Preliminary experiments over a 1.2 m link demonstrate effective transmission under strong turbulence ($D/r_0 \sim 5.0$) without requiring AO systems, paving a new direction for turbulence-resilient FSO architectures.

\section{Concept}
Unlike traditional FSO systems that encode data into the intensity or phase of a single source, GM utilizes the spatial geometric configuration of multiple sources. Fig.~\ref{fig:01}(a) illustrates the GM concept. The transmitter consists of a constantly-lit reference source $S_0$ and signal sources $S_1, \ldots, S_M$. In each symbol period, a specific subset of signal sources is activated to form different encoding patterns (Fig.~\ref{fig:01}(b)). While simultaneous activation can construct high-dimensional constellations, this proof-of-concept focuses on fundamental single-pair activations (Symbols 0, 1, 2).

The proposed active correlative decoding method contributes to turbulence resistance. It involves applying a frequency shift $f_0$ between $S_0$ and the signal sources. The receiver employs a detector array $D_0, D_1, \ldots, D_N$ to detect the correlation properties of the optical field in the overlapping region. The alternating current (AC) component $\hat{I}_n(t)$ is extracted via high-pass filtering. By multiplying each $\hat{I}_n(t)$ of detector with $\hat{I}_0(t)$ of the reference detector $D_0$ and time-averaging (low-pass filtering) over the symbol period $T$, we obtain the robust second-order spatial coherence:
\vspace{-2mm}
\begin{equation}
    \bar{g}^{(2)}(n) = \frac{\left\langle \hat{I}_n(t) \cdot \hat{I}_0(t) \right\rangle_T}{\sqrt{\left\langle \hat{I}_n^2(t) \right\rangle_T} \cdot \sqrt{\left\langle \hat{I}_0^2(t) \right\rangle_T}},
    \vspace{-2mm}
    \label{equ}
\end{equation}
where $\langle \cdot \rangle_T$ denotes the time average, and the denominator represents the product of the root-mean-square (RMS) values of the AC components.

Physically, the remarkable robustness of this method stems from the common-mode rejection of turbulence-induced perturbations. Let the AC component at the $n$-th detector be expressed as:
\vspace{-2mm}
\begin{equation}
    \hat{I}_n(t) = A_n(t) \cos(2\pi f_0 t + \Phi_{\text{geo}}(x_n) + \phi_{\text{turb}}(x_n, t)),
    \vspace{1mm}
\end{equation}
where $A_n(t)$ denotes the amplitude subject to turbulence-induced intensity scintillation, $\Phi_{\text{geo}}(x_n)$ is the deterministic geometric phase dictated by the source array, and $\phi_{\text{turb}}(x_n, t)$ is the random phase fluctuation introduced by the atmospheric channel. 

Crucially, because the spatial scale of the receiver array is relatively compact, the turbulence-induced phase fluctuations across the detectors are highly correlated, yielding $\phi_{\text{turb}}(x_n, t) \approx \phi_{\text{turb}}(x_0, t) \equiv \phi_{\text{turb}}(t)$. Consequently, the differential turbulence phase becomes purely time-dependent. When multiplying the signals from the $n$-th and the reference detector, we obtain a time-independent baseband term and a double-frequency term:
\vspace{-2mm}
\begin{equation}
    \begin{aligned}
        \hat{I}_n(t)& \cdot \hat{I}_0(t) = \frac{1}{2} A_n(t) A_0(t) \cdot \qquad \\
        \big\{ & \cos[\Phi_{\text{geo}}(x_n) - \Phi_{\text{geo}}(x_0)] \\
        & \qquad + \cos[4\pi f_0 t + \Sigma\Phi_{\text{geo}} + 2\phi_{\text{turb}}(t)] \big\}.
    \end{aligned}
    \vspace{-2mm}
\end{equation}
By performing the time average $\langle \cdot \rangle_T$ over the symbol period ($T \gg 1/f_0$), the high-frequency term—which contains the phase jitter $2\phi_{\text{turb}}(t)$—is canceled out. Furthermore, the random intensity scintillation effects embedded in $A_n(t)$ and $A_0(t)$ are effectively canceled out by the RMS normalization denominator and time-averaging in Eq.~\ref{equ}. 

Therefore, the system inherently resists both time-varying phase jitter and intensity scintillation, elegantly isolating the deterministic spatial correlation $\bar{g}^{(2)}(n) \propto \cos(k_m \cdot x_n)$ without requiring any adaptive optics, where $k_m=2\pi R_m/\lambda L$ is the spatial frequency determined by the spacing $R_m$ between $S_0$ and $S_m$, $\lambda$ is the wavelength, and $L$ is the propagation distance.

\section{Experimental Setup}
\begin{figure}[!bt]
    \centering
    \includegraphics[width=\columnwidth]{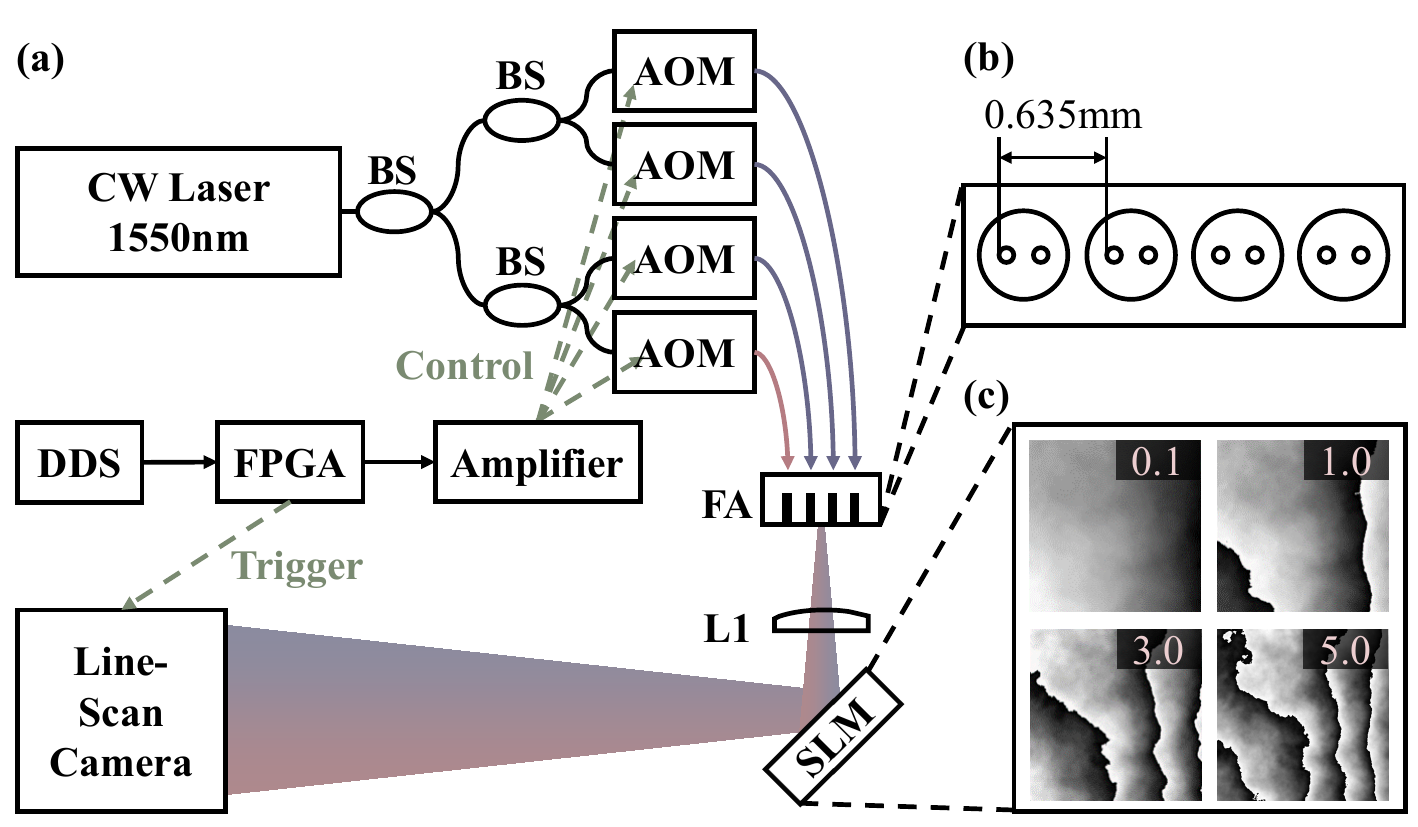}
    \caption{\textbf{(a)}~Proof-of-concept setup. Transmitter based on a horizontal fiber array driven by independent AOMs. Receiver composed with a horizontal line-scan camera. BS: beam splitter; FA: polarization-maintaining fiber array; L1: Cylindrical lens (only for shaping the vertical propagation of beam). \textbf{(b)}~Fiber array end-face. \textbf{(c)}~Simulated turbulence phase screens via SLM ($D/r_0 \in [0.05, 5.0]$).}
    \label{fig:02}
\end{figure}

The experimental setup is shown in Fig.~\ref{fig:02}(a). A 1550~nm continuous-wave laser is coupled into a 1D horizontal arranged polarization-maintaining fiber array (0.635~mm spacing, Fig.~\ref{fig:02}(b)) driven by four independent acousto-optic modulators (AOMs). An FPGA-DDS circuit synchronously controls the AOMs, setting a 80.002~MHz frequency shift to $S_0$ and 79.998~MHz to the others, establishing a frequency shift of $f_0 = 4$~kHz.

The optical field propagates over a 1.2 m free-space link. To reproducibly simulate dynamic atmospheric turbulence, a phase-only spatial light modulator (SLM) loads Kolmogorov phase screens \cite{RN16} statically. We selected 20 turbulence strengths within $D/r_0 \in [0.05, 5.0]$, generating 2000 independent phase screens per strength (Fig.~\ref{fig:02}(c)). At the receiver, a high-speed infrared line-scan camera (38.261~kHz, 512 pixels), parallel to the fiber array, captures the beat signal in the overlaping region. Each 10 samples of every pixel in time domain is processed via Eq.~\ref{equ}. The output $\bar{g}^{(2)}(n)$ is mapped to the spatial frequency domain via FFT, and decoded using max-peak position matching.

\section{Results}
We evaluated two-state (Symbol 0/1) and three-state (Symbol 0/1/2) schemes with four equally spaced sources. The system operates at a symbol rate of $\sim$3.8~kbaud/s, processing $3 \times 10^5$ symbols per turbulence strength.

\begin{figure*}[tb!]
    \centering
    \includegraphics[width=\textwidth]{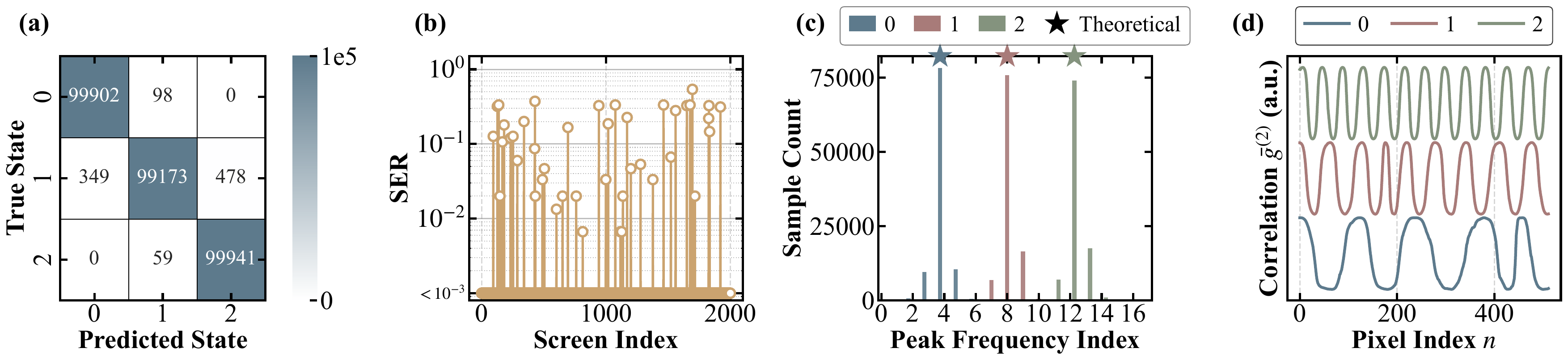}
    \caption{\textbf{System performance under strong turbulence ($D/r_0=5.0$).} \textbf{(a)}~Confusion matrix. \textbf{(b)}~SER across 2000 phase screens. \textbf{(c)}~Distribution of peak spatial frequencies. \textbf{(d)}~Normalized spatial second-order correlation $\bar{g}^{(2)}(n)$.}
    \label{fig:03}
\end{figure*}

Fig.~\ref{fig:03} details the three-state scheme performance under strong turbulence ($D/r_0 = 5.0$). The confusion matrix (Fig.~\ref{fig:03}(a)) indicates near-perfect classification despite wavefront distortion. Fig.~\ref{fig:03}(b) tracks the symbol error rate (SER) across 2000 independent phase screens, showing that errors, in most of the cases, are overwhelmingly suppressed below the measurement limit ($<10^{-3}$). This resistance stems from the fact that the extracted peak spatial frequencies converge tightly to theoretical positions without obvious cross-talk (Fig.~\ref{fig:03}(c)). As shown in Fig.~\ref{fig:03}(d), although in the interference of strong turbulence, the second-order spatial correlation $\bar{g}^{(2)}(n)$ dictated by the source geometry is still effectively preserved.

\begin{figure}[t!]
    \centering
    \includegraphics[width=\columnwidth]{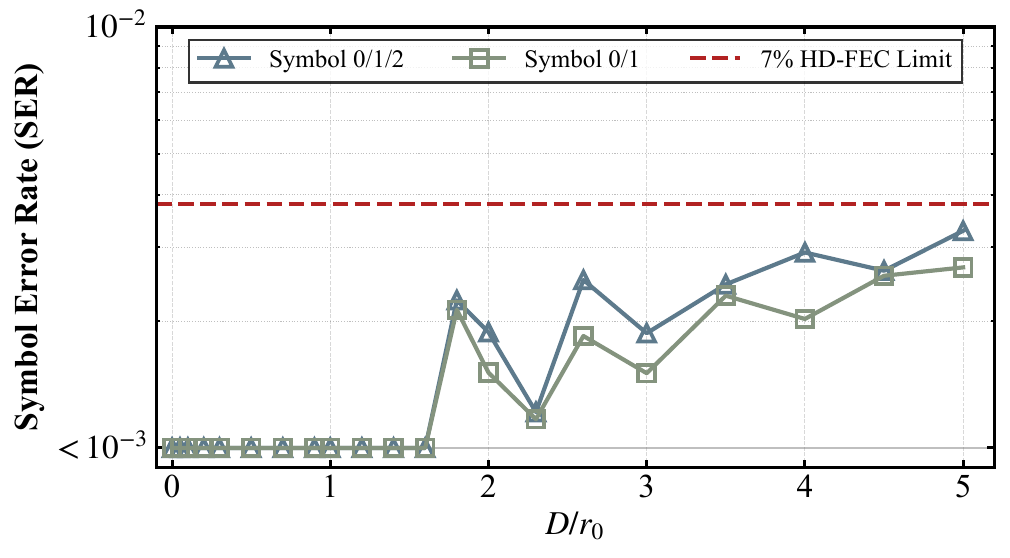}
    \caption{\textbf{SER vs. turbulence strength $D/r_0$ for two-state and three-state schemes.} Red dashed line denotes the 7\% HD-FEC limit.}
    \label{fig:04}
\end{figure}

Fig.~\ref{fig:04} summarizes the global SER performance. Both schemes exhibit remarkable robustness across the entire range. Even in the strong turbulence region ($D/r_0 = 5.0$), SER of GM strictly remains below the 7\% Hard-Decision Forward Error Correction (HD-FEC) threshold ($3.8 \times 10^{-3}$). 

\begin{table*}[b!]
    \centering
    \scriptsize 
    \caption{Comparison of State-of-the-Art FSO Modulation Paradigms under Atmospheric Turbulence}
    \label{tab:sota}
    \renewcommand{\arraystretch}{1.5} 
    \begin{tabular}{ >{\raggedright\arraybackslash}p{1.8cm} >{\raggedright\arraybackslash}p{2.6cm} >{\raggedright\arraybackslash}p{3.0cm} >{\raggedright\arraybackslash}p{2.9cm} >{\raggedright\arraybackslash}p{3.6cm} }
        \hline
        \textbf{} & \textbf{Conventional (OOK/QAM)} & \textbf{Orbital Angular Momentum (OAM)} & \textbf{Optical Spatial Modulation (OSM)} & \textbf{Geometric-configuration Modulation (GM, Ours)} \\
        \hline
        \textbf{Encoding Dimension} & Intensity / Phase (1st-order) & Helical wavefront (Spatial Phase) & Position of sources (Spatial Index) & \textbf{Geometric configuration (2nd-order Spatial Coherence)} \\
        
        \textbf{Turbulence Vulnerability} & Amplitude scintillation, rapid phase jitter & Modal crosstalk, topological charge degradation & Beam wandering & \textbf{Inherently immune to common-mode phase/intensity disturbances} \\
        
        \textbf{Mitigation Requirement} & Adaptive Optics (AO) for phase; dynamic thresholding & High-order AO combined with complex MIMO-DSP & Precise beam tracking, spatial diversity & \textbf{AO-Free; simple filter, FFT and matching} \\
        
        \textbf{Resistance} & Weak to Moderate & Moderate & Weak to Moderate & \textbf{Strong ($D/r_0 \ge 5.0$)} \\
        \hline
    \end{tabular}
\end{table*}

\section{Discussion}
Tab.~\ref{tab:sota} highlights the superiority of the proposed GM. By providing inherent resistance against phase and intensity disturbances, it facilitates robust transmission under strong turbulence conditions ($D/r_0 \ge 5.0$). This paradigm effectively eliminates the reliance on cost-intensive AO configurations, demonstrating that reliable FSO links can be established solely through simple filter, FFT and matching.

The current rate of 3.8~kbaud/s is strictly bottlenecked by the AOM acoustic transit time and camera frame rate, rather than fundamental GM limitations. Scaling to Gbps standards just requires hardware upgrades: replacing AOMs with GHz-bandwidth electro-optic modulators (EOMs) and upgrading the receiver to high speed photodetector arrays coupled with real-time digital signal processing (DSP). Furthermore, superimposing multiple source pairs and extending to a 2D planar array will exponentially expand the available geometric configurations, unlocking high-order spatial constellations (e.g., 256-GM) for high-capacity FSO links.

\section{Conclusions}
We proposed and experimentally demonstrated GM, a novel AO-free FSO paradigm based on second-order spatial coherence. With multi-source geometric configuration encoding and active correlative decoding, the system inherently resists atmospheric phase and intensity fluctuations, maintaining an SER below the HD-FEC limit even under strong turbulence. This robust and highly scalable architecture extends the physical-layer foundation for next-generation, high-reliability FSO networks.

\clearpage
\section{Acknowledgements}
The authors extend their sincerest appreciation to Prof. Bangfen Zhu for his exceptional guidance and continuous support. We also thank Chen Chen for his valuable contributions to the design and construction of the hardware circuits. This work was supported by the National Key Research and Development Program of China (Grant No. 2024YFA1210600).

\defbibnote{myprenote}{}
\printbibliography[prenote=myprenote]

\vspace{-4mm}

\end{document}